\documentclass[aps,prl,twocolumn,citeautoscript,superscriptaddress]{revtex4}
\usepackage{amsmath}
\usepackage{amssymb}
\usepackage{graphicx,bm}

\usepackage{lipsum}
\usepackage{hyperref}
\hypersetup{
	colorlinks=true,
	linkcolor=magenta,
	filecolor=violet,     
	urlcolor=blue,
	citecolor=red
}
\UseRawInputEncoding

\begin{document}

%\preprint{APS}

\title{L-MM Auger electron emission from chlorinated organic molecules under proton impact: angular distribution and total cross section measurment}
%/$\text{CCl_4,CH_2Cl_2,CHCl_3}$ and $C_6H_5Cl$}
%\\with Forced Linebreak}% Force line breaks with \\
%\thanks{Proton impact LMM Auger decay yield of Cl in CHCl$_3$ and C$_6$H$_5$Cl}%

\author{Rohit Tyagi}
\affiliation{Department of Physics, Indian Institute of Technology Kanpur, Kanpur - 208016, India}
\author{Abhijeet Bhogale}
\affiliation{Tata Institute of Fundamental Research, 1 Homi Bhabha Road, Colaba, Mumbai - 400005, India}
\author{Sandeep Bari}
\affiliation{Department of Physics, Indian Institute of Technology Kanpur, Kanpur - 208016, India}
\author{L. C. Tribedi}
\affiliation{Tata Institute of Fundamental Research, 1 Homi Bhabha Road, Colaba, Mumbai - 400005, India}
\author{A. H. Kelkar}
\email{akelkar@iitk.ac.in}
\affiliation{Department of Physics, Indian Institute of Technology Kanpur, Kanpur - 208016, India}


\date{\today}% It is always \today, today,

\begin{abstract}
We have measured absolute total cross section for LMM Auger electron emission of Cl in chlorinated methane and benzene chloride in collision with H$^+$ ion. Projectile energy dependence of the total yield as well as the angular distribution has been studied. Incident proton energy has been varied from 125 keV to 275 keV in steps of 50 keV. C KLL Auger yield have been compared with previous studies and found to be in agreement within the effect of chemical species It has been found that the LMM Auger yield of Cl is much more significantly affected by molecular environment than the C KLL. 

\end{abstract}

\maketitle

\section{Introduction}
The study of Auger electron emission in atoms has been extensively utilized in past several decades to investigate the electronic structure of atoms and molecules in gaseous as well as condensed phase. In additions, the technique is widely used in fields such as material science, metallurgy, and the microelectronics industry. In recent years, the study of fragmentation dynamics of molecules following inner shell ionization has opened a new avenue for studying molecular states of polyatomic molecules in unprecedented detail \cite{Stoychev2008, Trinter2014, lago2007, Kokkonen2018, Bolognesi_2020}. This has significant implications in atmospheric physics, biophysics, space science, and cometary science. Coincidence measurements of Auger electrons are crucial in determining the dissociation pathways following inner shell ionization. While the inner shell molecular orbitals closely resemble the atomic orbitals, the excitation and ionization spectra often exhibit notable differences. Previous studies have demonstrated variations of more than 30$\%$ in Auger electron yields owing to the molecular nature of the target \cite{Ariyasinghe1990, lapicki2005}. One possible explanation for this change in yield is the elastic scattering of the ejected Auger electrons from the surrounding atoms within the molecule \cite{Matthews1978}. The effect of chemical environment on Auger electron emission yields from molecules has been studied earlier \cite{Bhalla1987,MCDANIEL1987500} and a semi-empirical relationship between the number of surrounding atoms and the probability of scattering was formulated \cite{Matthews1978}. However, the semi-empirical theory was constructed for symmetric molecules where the Auger emission took place from the central atom. Furthermore, the role of higher order processes such as, double Auger decay, have not been considered in much detail. 

Chloromethanes which are part of the family of volatile organic compounds (VOCs) are suitable target systems for such studies. VOCs are organic chemicals that can easily evaporate into the air at room temperature. They play a significant role in the formation of ground-level ozone and secondary organic aerosols, which can have implications for air quality and climate. Chloroform (CHCl$_3$) is one example of a VOC that has shown a decline in global abundance. On the other hand, emissions of some other chlorinated methanes, such as dichloromethane (CH$_2$Cl$_2$) and chloroform (CHCl$_3$), have been increasing in recent years. These compounds are used as solvents and degreasers in various industries, including manufacturing, cleaning, and paint stripping. Their increased use and emissions have resulted in a significant rise in their tropospheric concentrations.

%Chloromethanes except carbon tetrachloride (CCl$_4$) are not regulated by either the Montreal Protocol or the Kyoto Protocol, an agreement that regulates the production of ozone-depleting compounds, due to their short half-life and predominantly natural origin. These short-lived Volatile organic compounds are therefore present in the atmosphere. Although it has been noted that the global abundance of chloroform has been declined but emissions of some chloro-methanes, such as CH$_2$Cl$_2$ and CHCl$_3$, have been grown in recent years due to their use as solvents and de-greasers in various industries leading to dramatically increase in their troposphere abundances. When a Chlorine (Cl) atom comes in contact with Ozone (O$_3$) molecule it can produce very efficiently chlorine monoxide (ClO) which after meeting another oxygen atom, breaks up and releases Cl atom which can destroy another molecule of Ozone forming a catalytic chain of Cl production and leading to damage up to 100,000 ozone molecules before removal from the stratosphere.

In recent years, there has been extensive research investigating the relationship between the valence shell structure and fragmentation of chlorinated methane molecules using UV and VUV excitation \cite{Kokkonen2018}. Electron ion coincidence techniques have been employed to study the dissociation dynamics of CCl$_4$, CHCl$_3$, CH$_2$Cl$_2$\cite{Lu_2008, Lu2011, Alcantara2016} upon near 2p shell excitation. A significant fragmentation pathway that has been observed involves Auger electron emission following the 2p vacancy of the Cl atom in these molecules, resulting in liberation of chlorine atoms. 

In the present work, we have measured the absolute total cross section of LMM Auger electron emission from chlorinated molecules in collisions with 100 keV to 275 keV proton. Previous literature has predominantly focused on investigating the influence of electronegative atoms (specifically halogens) on the Auger yield of carbon (C) and sulfur (S). Our work expands upon this by examining the impact of both carbon and chlorine (Cl) on the Auger emission yield of chlorine. In this study, we also measured the angular distribution of the emitted electrons.

\section{Experimental details}
The experimental measurements were performed at the 300 keV ECR ion accelerator (ECRIA) facility at TIFR Mumbai \cite{MANDAL201919,Agnihotri_2011}. A beam of H$^+$ ions was obtained from the ECR ion Source floated at 30 keV. The ECR ion source itself is mounted on a 300 keV high voltage (HV) deck and the HV deck was raised to appropriate potential in order to obtain ion beams with desired energy. Post acceleration the projectile ion beam was diverted towards the main scattering chamber connected to the 30$^o$-north beam line. The projectile ion beam was collimated and focused using an electrostatic quadrupole lens riplet and a pair of four jaw slits. A differential pumping chamber with a pair of entrance (2 mm) and exit (4 mm) collimators was connected before the main scattering chamber to obtain near parallel projectile ion beam for collisions. The collimators also helped in reducing the ion beam induced low energy electron background during the experiments. 
%Experiments have been performed in two separate beam time with ECR ion source Accelerator facility at TIFR, Mumbai\cite{MANDAL201919,Agnihotri_2011}.   During both the beam time experimental conditions have been kept same and proton beam was provided by 14.5 GHz cyclotron frequency 400-kV accelerator facility, in which plasma was created with hydrogen gas and extraction voltage was kept 30kV. A switching magnet is utilised to direct the beam toward the selected beam-line out of four beam-line. For focusing and guiding the ion beam,  beam-line have  quadruple lenses and X-Y deflectors. Additionally, the beam cross-section and the beam divergence is controlled by a pair of four-jawed slits. To maintain the beam-line vacuum and to cut the broad beam, a chamber with and extended collimator with diameters of 2 mm at the entrance and 4 mm at the exit was pumped deferentially.
Energy selection and detection of the secondary electrons post collision was done using a electron spectrometer, which is an electrostatic hemispherical analyzer mounted on a rotating turn table inside the main scattering chamber. The electron spectrometer has $\sim$ 6$\%$ resolution. Complete details of the experimental setup and spectrometer performance have been described earlier \cite{MISRA2009}. The measurements were performed in flooded chamber mode in order to maintain uniform target pressure at all solid angles. The chloromethane targets are low vapor pressure liquids at room temperature, and convert to gaseous state when exposed to the high vacuum ($\sim$ 10$^{-8}$ mbar) of the main scattering chamber. The pressure in the scattering chamber was maintained below 5$\times$ 10$^{-5}$ mbar during the experiment to maintain single collision conditions.

The relative DDCS (double differential cross section) is calculated using the following relation:
\begin{equation}
    \frac{d^2\sigma}{d\Omega d\epsilon} = \frac{(\frac{n_e}{N_p\Delta\epsilon} - \frac{n_b}{N_p\Delta\epsilon})}{P_t\epsilon_{el}(l\Omega)_{eff}}
\end{equation}
Here, n$_e$ and n$_b$ are detected electron counts from the target gas and background respectively, N$_p$ is the number of projectile ions, P$_t$ is the target gas pressure, $\Delta\epsilon$ is the energy resolution of the emitted electrons, $\epsilon_{el}$ is the detection efficiency and $(l\Omega)_{eff}$ is the effective solid angle path length \cite{MISRA2009}.
%In brief, a hemispherical analyzer which has the entrance and exit slit dimension 3 $\times4$ mm$^2$ and 4$\times4$mm$^2$ respectively and typical resolution of $6\%$ was used for measurements.  More Details of the spectrometer are given in the reference\cite{MISRA2009}. Initially, stain less steel scattering chamber was evacuated to the typical base pressure of 1.1$\times10^{-8}$ mbar.Then the chamber was flooded with the vapour of different target gases, pressure of 0.6-1.3 mbar was maintained using a solenoid valve. 

The measured relative DDCS were then converted to absolute DDCS using suitable normalization with the existing data. For this normalization we measured the K-LL Auger spectra of Ne (atomic target) and CH$_4$ (molecular target). The Ne K-LL Auger spectrum was measured at few angles in collisions with 250 keV protons. By integrating the relative DDCS over the entire energy  and angular range we obtained the total K-LL Auger electron emission cross section which is normalized to the experimental cross sections measured by Stolterfoht et al. \cite{Stolterfoht1974}. A normalization factor of $\sim$ 1.73 was obtained. This normalization factor was then used to obtain the absolute total Auger cross section for methane target under identical experimental conditions for 100 keV proton beam. Our measured value of $\sim$ 1.53$\times$ 10$^{-19}$ cm$^2$ is in excellent agreement with the previously reported cross section values \cite{Stolterfoht1975}. Uncertainty in the measurement of the target gas pressure ($\sim$ 10$\%$) is the dominant source of error in the calculation of absolute DDCS, other than the statistical error. The over all uncertainty in absolute cross section is estimated to be $\sim$ 15$\%$.

As a systematic study, we have measured the DDCS for chlorine containing organic targets namely, CCl$_4$, CHCl$_3$, CH$_2$Cl$_2$, and C$_6$H$_5$Cl. The DDCS spectra were measured in two beam times spread over several days. The measurements were performed at backward angles of emission ranging from 90$^o$ to 150$^o$ in steps of 15$^o$ and at different projectile energies between 100 keV to 275 keV. 
\begin{figure}
\includegraphics[width = 0.45\textwidth]{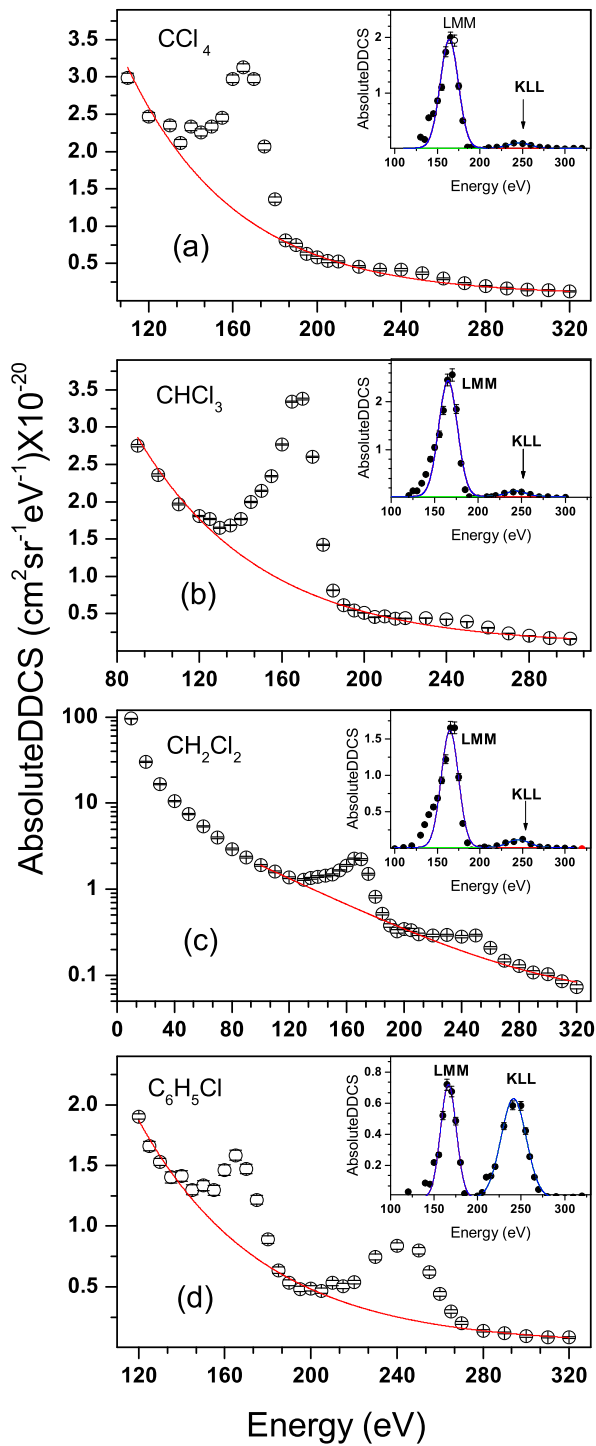}% Here is how to import jpg art
\caption{Absolute DDCS electron emission spectra on bombarding H$^+$ ion of energy (a) 200 keV (b) 275 keV (c) 250 keV (d) 225 keV for CCl$_4$,CHCl$_3$ CH$_2$Cl$_2$ and  C$_6$H$_5$ molecule respectively. In captions LMM Auger peak of Cl and KLL of carbon after continuum background subtraction are shown.}
\label{fig1}
\end{figure}

\section{Results and discussion}

%\subsubsection{ Double differential cross section - DDCS}

 Figure \ref{fig1} shows the measured absolute DDCS spectra for CCl$_4$, CHCl$_3$, CH$_2$Cl$_2$, and C$_6$H$_5$Cl molecular targets. The spectra were taken for secondary electron energy range 100 eV to 300 eV covering the Cl L-MM and C K-LL Auger energy range. In the present study we have restricted the measurements in the Auger energy range (high energy electrons), however, complete electron spectrum down to 10 eV was taken in a few cases to verify the spectrometer performance over the entire energy range. For example, complete electron spectrum for CH$_2$Cl$_2$ molecular target is shown in figure \ref{fig1}c. The shape of the spectrum is charactestic of secondary electron emission due to binary collision. In the low energy region the cross section is dominated by soft electron emission and as the emission energy increases, the cross section of this continuum electron background decreases, monotonically, by several orders of magnitude. The DDCS spectra in figure \ref{fig1} show that the Auger peaks corresponding to Cl L-MM ($\sim$ 170 eV) and C K-LL ($\sim$ 240 eV) are visible over the continuum background electron spectrum. The angular distribution of the continuum electrons has been studied in detail for various targets \cite{MISRA2009, Bagdia2022, Kelkar2020, Bhattacharjee2016} and it has been established that the emission cross sections are larger at forward scattering angle compared to backward scattering angles. In our measurements with chloromethanes, we observed that the continuum electron background dominated the electron spectrum at forward angles such that it was not possible to analyse the contribution from the Auger peaks. Therefore, we have restricted our measurements for backward emission angles (90$^o$ to 150$^o$). 

 In order to extract the contribution from Auger peaks, one can approximate the continuum background in the peak region by a polynomial function and subtract it from the measure DDCS spectrum. This interpolated continuum background is shown as solid curve in figure \ref{fig1} and the subtracted Auger peaks are shown in the inset. It is also observed that in the DDCS spectrum for C$_6$H$_5$Cl (see figure \ref{fig1}d), the C K-LL peak is notably more prominent compared to the spectra of CCl$_4$, CHCl$_3$, and CH$_2$Cl$_2$ (figure \ref{fig1}(a-c)). The relatively larger C K-LL auger yield in C$_6$H$_5$Cl is due to the higher number of C-atoms present in the aromatic molecule.

The peak corresponding to Cl L-MM Auger emission (centered at $\sim$ 170 eV) in all the spectra results from a combination of both atomic transitions (resulting in distinct, sharp lines) and molecular transitions (resulting in broader lines due to varying inter-nuclear distances) \cite{Santos2018}. With a spectrometer resolution of $\sim$ 6$\%$, it is not possible to separate the individual yields due to different transitions. However, the total yield, encompassing both atomic and molecular contributions, can be determined from the spectrum. The Cl L-MM peak spans an energy range 130 eV to 200 eV. The peak has contributions from three primary transitions of Auger electron energy occurring at 150 eV, 165 eV, and 170 eV \cite{Ariyasinghe1990-1}.

\begin{figure}
\includegraphics[width = 0.45\textwidth]{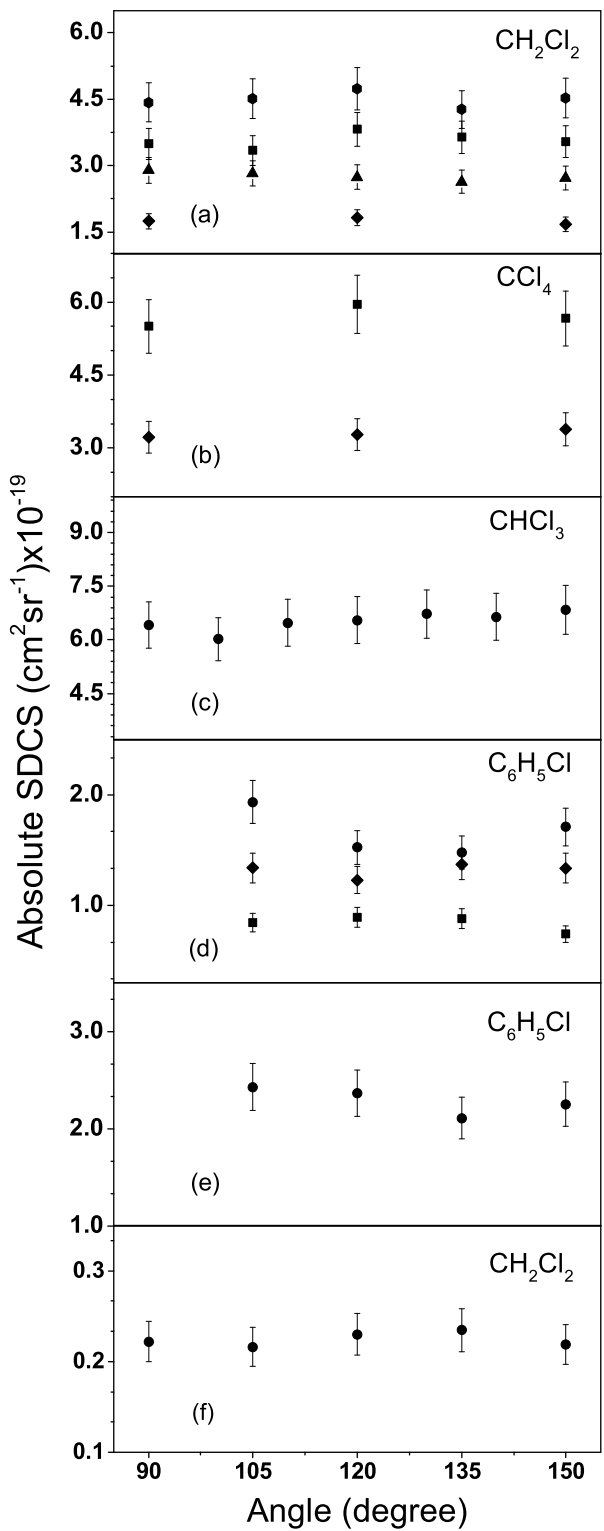}
\caption{ Angular distribution of Absolute SDCS with respect to the projectile beam for Cl LMM (a)CH$_2$Cl$_2$ (250keV, 200keV, 150 keV and 100 keV)  (b)CCl$_4$ (100 keV , 200 keV) (c)CHCl$_3$ (d)C$_6 $H$_5$Cl (125,175,225 keV) and C KLL (e)C$_6 $H$_5$Cl (225 keV) (f)CH$_2$Cl$_2$ (250 keV).}
\label{fig2}
\end{figure}

The angular distribution of the Auger electrons carries information about the electron density distribution of the interacting orbitals. For a multi atomic target, the angular distribution may also reveal the effects of molecular environment on the Auger emission process. The angular distribution has been studied in quite extensively  for K-auger decay from atomic and small molecular targets upon electron and ion collision. In all cases the K-auger emission was found to be isotropic with respect to the projectile beam direction \cite{Matthews1978, Stolterfoht1974, Toburen1972, Kelkar2022}. This is mainly due to the spherical symmetry of the K-shell involved. On the other hand, it was proposed that for Auger decays involving states with total angular momentum j $> \frac{1}{2}$, the angular distribution of the emitted electrons could be anisotropic if the resulting atomic state has a quantum number j$\:' > \frac{1}{2}$ \cite{flugge1972, mehlhorn1968}, which was verified by subsequent measurements of Ar L-MM Auger spectra angular disrtbution \cite{CLEFF19713, Paripas2004}. Auger electron emission from molecules are still different from the atomic case and existence of various excitation levels with different symmetry and shape resonances in continuum may contribute towards enhancing or suppressing the anisotropy\cite{Lindle1984,Becker1986,Hemmers1993,Edwards1997,HILTUNEN1999,Colle2004,Kadar2004}. The anisotropy in Ar L-MM Auger yield was found to be largest ($\sim$ 10$\%$) when the projectile electron collision energy was three times the L$_3$ subshell binding energy. For the case of heavy ion collision it is proposed that the angular anisotropy in Auger emission will be maximum when the projectile ion velocity is equal to the velocity of the bound electron in the L-shell.

In figure \ref{fig2} we have plotted the angular distribution of L-MM Auger emission from various chloromethane targets in collisions with keV energy proton beam. The projectile velocity in our experimental energy range (100 keV  - 275 keV) is varies from $\sim$ 0.9$v_e$ to $\sim$ 0.5$v_e$, where $v_e$ is the velocity of bound electrons in Cl L-shell. It is evident from the plots in figure \ref{fig2} that the Cl L-MM Auger angular distribution is isotropic, within the experimental resolution, for all projectile energy and target combinations. A similar isotropic angular distribution was also reported earlier \cite{Guo1995}, for 1.2 MeV and 1.8 MeV He$^+$ projectile ions. As mentioned in the text above, the spectrometer resolution in present experiment is $\sim$ 6$\%$, which is not sufficient to identify subshell resolved contribution to the total Auger yield. Nevertheless, the observed isotropic angular distribution signifies the dominant contribution of the final state Cl$^{2+}$[3p$^{-2}$] term $^2$P$_{1/2}$, with j $= \frac{1}{2}$ \cite{Hrast2019}. Furthermore, the isotropic angular distribution plots for all the four target molecules (see figure \ref{fig2}) show that the molecular nature of the target has very feeble effect, if any, on the angular distribution and the relative excitation probability of the transition states contributing to the Auger process does not depend strongly on the molecular nature of the target. In figure \ref{fig2}e we have also shown the C K-LL angular distribution measured for CH$_2$Cl$_2$ molecule along with the Cl L-MM auger emission distribution. The isotropic nature of the C K-LL angular spectrum, as already established by several experimental investigations \cite{Kelkar2022} (and references therein,) confirms that the nature of Cl L-MM angular distribution is devoid of any experimental artifact, which may smear out probable anisotropy in the measured data. 

It is also interesting to note here that although the molecular environment should not affect the core excited states, as expected, however it may lead to observable changes in energy of the ejected electrons and their angular distribution (in molecular frame) due to rescattering in the outgoing channel. The change in ejected electron energy spectrum has been observed as broadening of the Auger peaks in molecular targets \cite{Santos2018}. To investigate the effect of rescattering from the surrounding atoms of the molecules orientation of the target molecules needs to be determined. In the present experiment we can not investigate this effect since the target molecules are oriented isotropically relative to the projectile beam direction. However, electron-ion coincidence measurements offer such a possibility \cite{Guillemin2015}. 

\begin{figure}
\includegraphics[width = 0.45\textwidth]{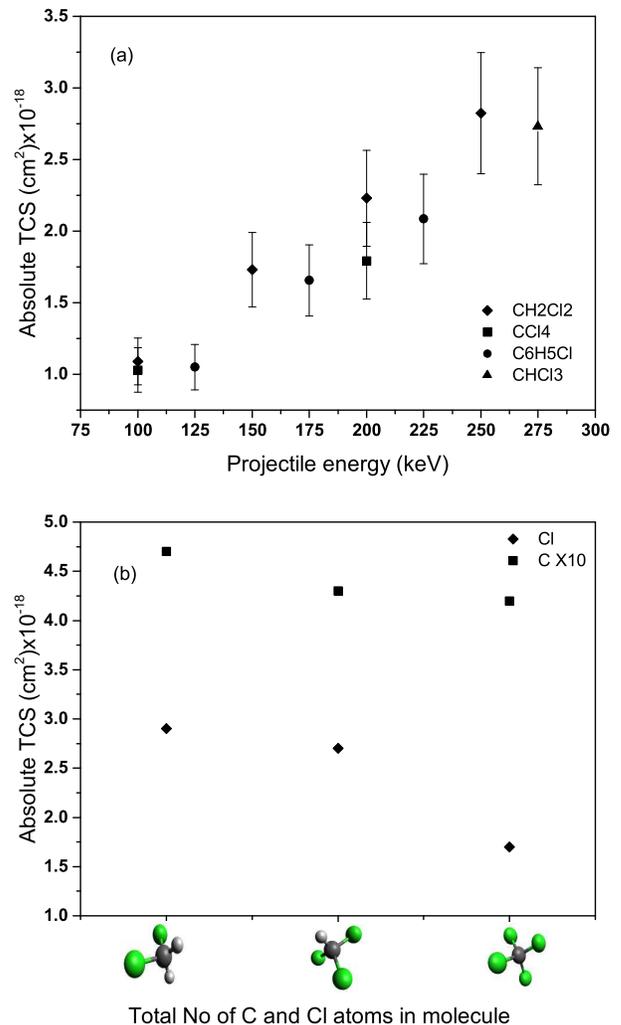}
\caption{Total cross section of LMM Auger transition of Cl for different projectile energy in chlorinated methane and C$_6$H$_5$Cl.}
\label{fig3}
\end{figure}

The total absolute Cl L-MM Auger yield was calculated from the measured DDCS by taking the mean cross section from the angular distribution plot (see figure \ref{fig2}) and integrating it over 4$\pi$ solid angle. The values are listed in table \ref{table1} We have also listed the total cross section values for C K-LL Auger emission measured in each case from the same molecule. The Cl L-MM Auger emission cross section from various from various chlorine containing molecules have been reported earlier for collisions with MeV energy He$^+$ ions \cite{Ariyasinghe1990}. In the present projectile energy range, previous data is available only for He$^+$ projectile beam. For comparison we have also included the cross section values measured by Ariyasinghe et. al \cite{Ariyasinghe1990} in table \ref{table1} for He projectile with the same energy/amu. This comparison (see table \ref{table1}) shows that the cross section for H$^+$ projectile is $\sim$ 4 times smaller than He$^+$ projectile. A similar ratio was earlier observed in the case of C K-LL Auger cross section \cite{Stolterfoht1974}. In fact, the C K-LL auger cross section measured in the present experiment are also in excellent agreement with those reported in \cite{Stolterfoht1974}. A careful look at C K-LL Auger cross section reported in table \ref{table1} also shows that the cross section at $\sim$ 100 keV projectile energy is rather low for CH$_2$Cl$_2$ molecule (0.9 $\times$ 10$^{-19}$ cm$^2$) compared to that for C$_6$H$_5$Cl molecule (1.7 $\times$ 10$^{-19}$ cm$^2$). In fact, for CCl$_4$ molecule the Auger peak corresponding to C K-LL is barely visible above the continuum electron background in the ddcs plot (at 100 keV projectile energy) and hence the total cross section could not be estimated. This behavior is due to the dominant contribution of continuum electrons in the C K-LL energy region from multiple Cl atoms in the parent molecules. In figure \ref{fig3}a, we have plotted the total Cl L-MM Auger yield (from all the target molecules) as a function of projectile energy. It is evident that the data shows a monotonically increasing behavior and even the cross section values from different molecules match within the experimental uncertainty. This is expected in the present energy range as predicted by theory and experiments \cite{Ariyasinghe1990, brandt1981, Hansen1973}. We also observe that cross section values from different molecules for the same projectile energy match within the experimental uncertainty. 

The variation in the measured cross section for different molecular targets is also attributed to the molecular environment surrounding the atom undergoing Auger decay. The presence of surrounding atoms in close vicinity (1-2 \AA) may cause rescattering of the Auger electron, thereby changing its energy. This results in reduction in the measured cross section. The rescattering probability due to presence of electronegative atoms was formulated in terms of a geometrical  expression \cite{Matthews1978} to estimate the change in total Auger emission cross section in terms of transmission loss. For the molecules under consideration (e.g. CCl$_4$ and CH$_2$Cl$_2$ ), a change in cross section $\sim$ 18$\%$ is estimated which is quite close to the measured difference as shown in table \ref{table1}. The effect of rescattering will also depend on the energy of the Auger electrons. This is shown in figure \ref{fig3}b, where we have plotted the total Auger cross section from Cl (L-MM) and C (K-LL) as a function of  total number of atoms in the target molecule. The plot shows that the cross section decreases as more number of atoms surrounding the Auger atom increases. We also see that the loss in cross section for Cl is steeper than that for C atom. This could be attributed to the higher energy of C K-LL Auger electrons compared to Cl L-MM Auger electrons.  

\begin{table}[]
\resizebox{0.9\columnwidth}{!}{
\begin{tabular}{|ccc|c|}
\hline
\multicolumn{1}{|c|}{\begin{tabular}[c]{@{}c@{}}ion energy\\  (keV/amu)\end{tabular}} &
  \multicolumn{1}{c|}{\begin{tabular}[c]{@{}c@{}}Cl\\ (10$^{−18}$ cm$^2$)\end{tabular}}&
  \begin{tabular}[c]{@{}c@{}}C \\ (10$^{−19}$cm$^2$)\end{tabular} &
  \begin{tabular}[c]{@{}c@{}}Cl Cross Section Ref\cite{Ariyasinghe1990}\\ (10$^{-18}$cm$^2$) \\CH$_3$Cl target and He$^+$ projectile\end{tabular} \\ \hline
\multicolumn{3}{|l|}{CH$_2$Cl$_2$}                               & \multicolumn{1}{l|}{} \\ \hline
\multicolumn{1}{|c|}{100} & \multicolumn{1}{c|}{1.1} & 0.9 &                       \\ \hline
\multicolumn{1}{|c|}{150} & \multicolumn{1}{c|}{1.7} & 2.9 & 5.5                   \\ \hline
\multicolumn{1}{|c|}{200} & \multicolumn{1}{c|}{2.2} & 4.6 & 8.8                   \\ \hline
\multicolumn{1}{|c|}{250} & \multicolumn{1}{c|}{2.9} & 4.7 & 10.2                  \\ \hline
\multicolumn{3}{|l|}{CHCl$_3$}                                & \multicolumn{1}{l|}{} \\ \hline
\multicolumn{1}{|c|}{275} & \multicolumn{1}{c|}{2.7} & 4.3 &                       \\ \hline
\multicolumn{3}{|l|}{CCl$_4$}                                 & \multicolumn{1}{l|}{} \\ \hline
\multicolumn{1}{|c|}{100} & \multicolumn{1}{c|}{1.0} &     &                       \\ \hline
\multicolumn{1}{|c|}{200} & \multicolumn{1}{c|}{1.7} & 4.2 &                       \\ \hline
\multicolumn{3}{|l|}{C$_6$H$_5$Cl}                               & \multicolumn{1}{l|}{} \\ \hline
\multicolumn{1}{|c|}{125} & \multicolumn{1}{c|}{1.0} & 1.7 &                       \\ \hline
\multicolumn{1}{|c|}{175} & \multicolumn{1}{c|}{1.6} & 3.2 & 6.7                   \\ \hline
\multicolumn{1}{|c|}{225} & \multicolumn{1}{c|}{2.0} & 3.6 &                       \\ \hline
\end{tabular}
}

\caption{Absolute total cross section, C K-LL and Cl L-MM}
\label{table1}
\end{table}

\section{conclusion}
Total yeild of LMM Auger electron emission for Cl in CH$_2$Cl$_2$,CHCl$_3$,CCl$_4$ and C$_6$H$_5$Cl molecules, has been obtained under the impact of  H$^+$ ion of energy 100-275 keV. Angular distribution for the same has also been studied and found to be isotropic for these molecule within the experimental error. It is found that effect of chemical environment on C (KLL ) Auger yield is not much significant than the Cl (LMM). It also support the idea of yield loss in molecules is mainly due to the in-elastic scattering of Auger electron by valance shell electrons. Other processes which can affect the Auger electron yield are collective relaxation in which Double Auger decay is Dominant but quantification of that goes beyond the scope of this paper.

\section{Acknowledgments}
The authors would like to thank Mr. K. V. Thulasiram, Mr. Nilesh Mahtre and Dr. Deepankar Misra for facilitating the operation of the ECR ion accelerator, TIFR, Mumbai. 

%\nocite{*}
\bibliographystyle{utphys}
\bibliography{apssamp}% Produces the bibliography via BibTeX.
\end{document}